# Current-Induced Magnetization Switching in Permalloy-based Nanopillars with Cu, Ag, and Au.

H. Kurt, R. Loloee, W. P. Pratt Jr., and J. Bass
*Department of Physics, Center for Sensor Materials, and Center for Fundamental Materials Research, Michigan State University, East Lansing, MI 48824-2320*

## Abstract

We compare magnetoresistances (MR) and switching currents ($I_s$) at room temperature (295K) and 4.2K for Permalloy/N/Permalloy nanopillars undergoing current-induced magnetization switching (CIMS), with non-magnetic metals N = Cu, Ag, and Au. The N-metal thickness is held fixed at 10 nm. Any systematic differences in MR and $I_s$ for the different N-metals are modest, suggesting that Ag and Au represent potentially viable alternatives for CIMS studies and devices to the more widely used Cu.

Most experimental studies of current-induced magnetization switching (CIMS) in ferromagnetic/non-magnetic/ferromagnetic (F/N/F) trilayer metal nanopillars have used N = Cu as the spacer layer [see, e.g., 1-8]. None has yet used N = Ag or Au. Ag has the potential advantage for devices of sometimes giving a larger Current-Perpendicular-to-Plane (CPP) magnetoresistance (MR) with Permalloy (Py = Ni(84)Fe(16)) [9]. Au has the advantage of being insensitive to atmospheric contamination. We, thus, decided to compare MR and CIMS data at 295K and 4.2K for Py-based nanopillars with N = Cu, Ag, and Au.

Our sputtered samples have approximately elliptical shape with dimensions ~ 70 nm x 130 nm. dV/dI was measured with a lock-in amplifier at frequency ~ 8 kHz and measuring current ~ 20 μA. Details of sample preparation and measurements are given in [6,10]. Fig. 1 compares representative MR and CIMS switching data at 295K for samples of Py(24 nm)/N(10 nm)/Py(6 nm) nanopillars with N = Cu, Ag, and Au. Fig. 2 compares the same quantities for the same samples at 4.2K. Table I compares average values of the total resistance, R, the change in resistance upon switching, ΔR, the magnetoresistance, MR(%) = (ΔR/R)x100%, the difference $\Delta I_s = I_s^+ - I_s^-$ between positive (+ ) and negative (-) switching currents, $I_s$, and the upper magnetic switching field, $H_s$, over 3 to 5 samples of each type at 295K (top three data sets) and 4.2K (bottom three data sets). The average resistances of the three sets of samples differ, due to an unknown mix of variation in areas (smaller area increases R, does not change MR, and decreases $\Delta I_s$) and contact resistances (larger contact resistance increases R, decreases MR, and may leave $I_s$ unchanged). Thus, a precise comparison between them cannot be made. However, at both temperatures, the MRs and $\Delta I_s$ are roughly similar for all three metals. We conclude that Ag and Au represent potentially viable alternatives to Cu for studies of CIMS physics and for CIMS-based devices.

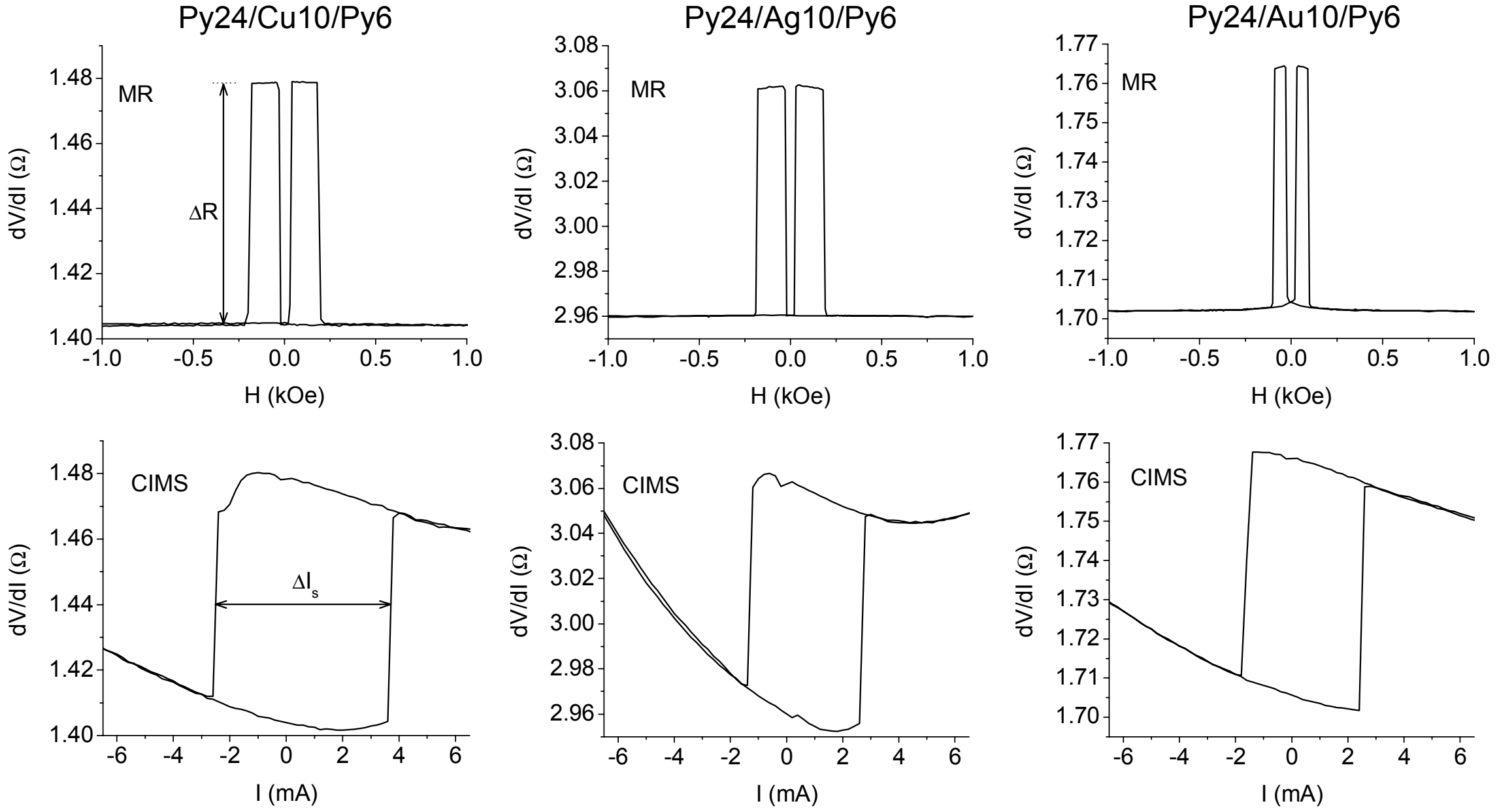

Fig. 1. MR(top) and CIMS(bottom) at 295K for Py/N/Py nanopillars with N = Cu (left), Ag (middle) and Au (right). Layer thicknesses are in nm.

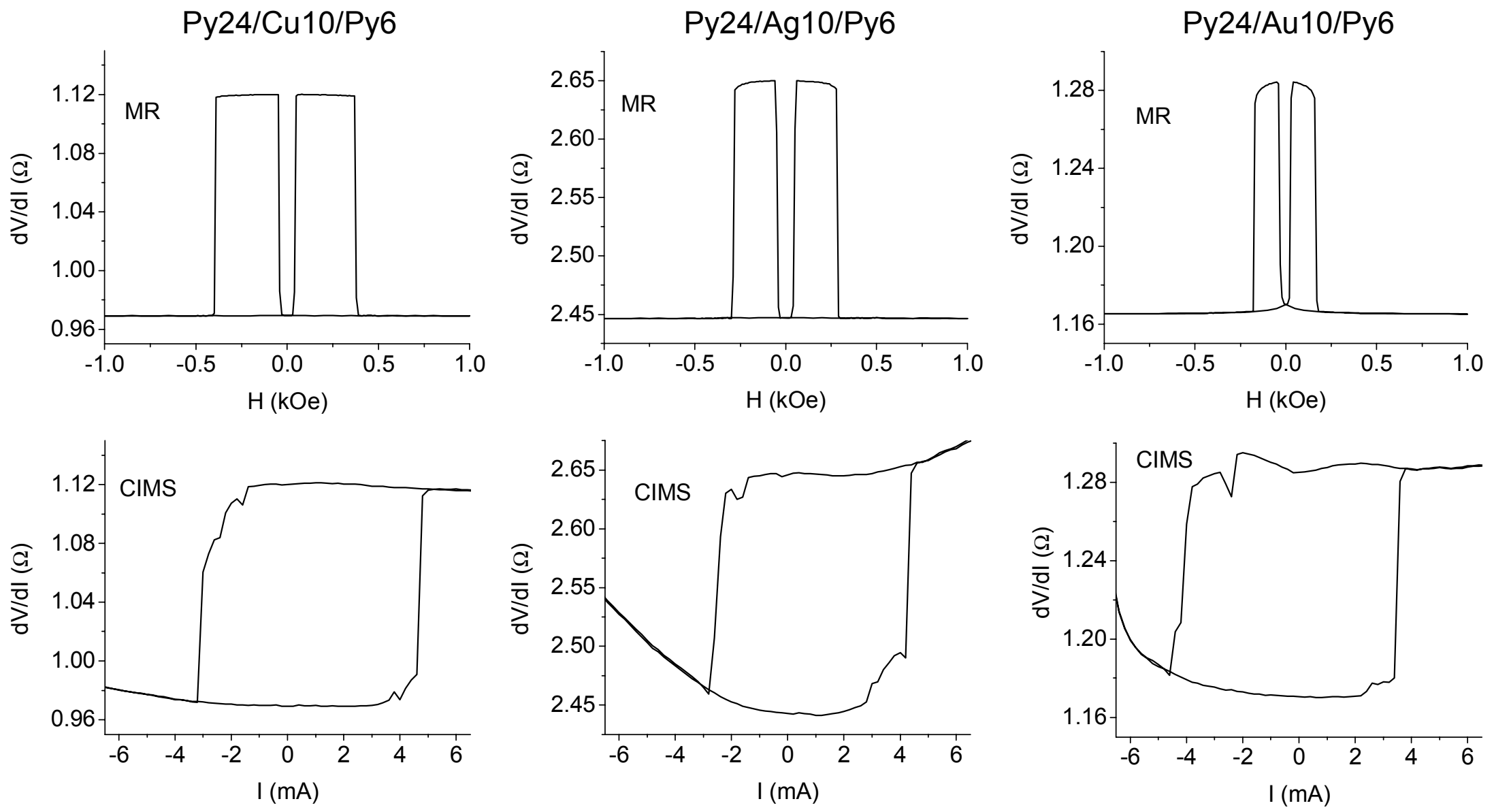

Fig. 2. MR (top) and CIMS (bottom) at 4.2K for Py/N/Py nanopillars with N = Cu (left), Ag (middle), and Au (right). Layer thicknesses are in nm.

| N-Metal | <R> (Ω) | <ΔR> (Ω) | <MR> (%) | <$\Delta I_s$> (mA) | <$H_s$> (kOe) |
|---|---|---|---|---|---|
| **295K** | | | | | |
| Cu | 1.4 | 0.07 | 5 | 5.7 | 0.2 |
| Ag | 2.6 | 0.09 | 3.5 | 3.2 | 0.14 |
| Au | 1.6 | 0.06 | 3.5 | 4.4 | 0.12 |
| **4.2K** | | | | | |
| Cu | 0.95 | 0.14 | 15 | 7.7 | 0.35 |
| Ag | 2.3 | 0.19 | 8.4 | 7.2 | 0.2 |
| Au | 1.3 | 0.12 | 9.4 | 7.5 | 0.18 |

Table I. Average values of R, ΔR, MR(%), $\Delta I_s$, and $H_s$ at 295K (top three rows) and 4.2K (bottom three rows) of Py/N/Py with N = Ag, Au, Cu. The 295K data are averaged for each metal over 3-5 samples; those at 4.2K are averaged over 3 samples.

Supported in part by the MSU CSM and CFMR, the US NSF under grants DMR 02-02476, 98-09688, NSF-EU collaborative grant 00-98803, and Seagate Technology.

References.